\documentclass[aps, pra, lengthcheck,letterm,nofootinbib]{revtex4-1}
\usepackage{amsfonts, amssymb, amsmath, graphicx, comment, bm, slashed}

\usepackage{dcolumn}   
\usepackage{bm}        
\usepackage{subfig}
\usepackage{caption}
\usepackage{multirow}

\usepackage{dcolumn}

\newcommand{\beq}{\begin{eqnarray}}
\newcommand{\eeq}{\end{eqnarray}}
\newcommand{\beqa}{\begin{eqnarray}}
\newcommand{\eeqa}{\end{eqnarray}}
\newcommand{\Lc}{{\cal{L}}}
\newcommand{\Uc}{{\cal{U}}}
\newcommand{\tr}{\mbox{tr}}

\def\braket#1{\mathinner{\langle{#1}\rangle}}

\begin{document}

\title{The axial anomaly and three-flavor NJL model with confinement : constructing the QCD phase diagram}

\author{Philip D. Powell}
\author{Gordon Baym}

\affiliation{Department of Physics, University of Illinois at Urbana-Champaign, 1110 W. Green Street, Urbana, Illinois 61801, USA}

\date{\today}

\begin{abstract}

We investigate the phase structure of massless three-flavor QCD by extending the Nambu--Jona-Lasinio model to include the effects of confinement and the axial anomaly.  We study the interplay between the chiral and diquark condensates induced by the axial anomaly, as well as their relationship with the Polyakov loop, which parameterizes confinement.  By minimizing the thermodynamic potential we construct the QCD phase diagram and investigate the possibility of realizing a recently discovered low temperature critical point and an associated BEC-BCS crossover.  We also perform a Ginzburg-Landau expansion of the thermodynamic potential, comparing our results to a prior analysis based purely on symmetry considerations, in order to assess the lowest-order effects of the condensate-confinement couplings.

\end{abstract}

\maketitle

\section{Introduction}

The phase structure of strongly-interacting matter has been a subject of great interest since the emergence of quantum chromodynamics (QCD) in the early 1970s.  However, it has gained greater attention in recent years as the boundaries of experimental study have continued to expand, particularly at the Relativistic Heavy Ion Collider (RHIC), and even now at the Large Hadron Collider (LHC).  In addition, more powerful computational techniques continue to produce more reliable lattice data, which may be compared with experimental results.

However, there remains much wanting in the theoretical analysis of QCD matter, due to the dual problem of the intractibility of nonperturbative QCD and the fermion sign problem, which limits our ability to extend lattice techniques to non-zero chemical potential.  One tool which has proven useful in filling this gap is symmetry-based models, which have been developed to describe critical characteristics of strongly-interacting matter including chiral symmetry breaking, diquark pairing, and confinement, while still retaining calculational tractibility~\cite{NJL1,NJL2,Pisarski,Volkov}.  Recently, increased attention has been given to the importance of the instanton-induced axial anomaly and the attractive chiral-diquark condensate coupling that it mediates~\cite{Kobayashi,tHooft,Gross,Boomsma}.

\begin{figure}
\includegraphics[scale=0.7]{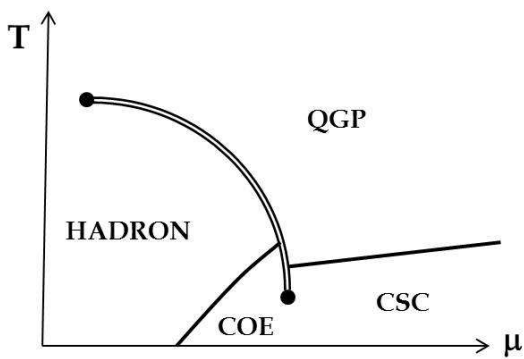}
\caption{\footnotesize{Schematic QCD phase diagram showing regions of broken chiral symmetry (hadron), BCS quark pairing and color superconductivity (CSC), coexistence of chiral and diquark condensates (COE), and a deconfined quark-gluon plasma (QGP).  Single lines denote second order phase transitions while double lines denote first-order transitions.  Adapted from~\cite{Yamamoto}.}}
\label{fig:schematic}
\end{figure}

The axial anomaly, which gives rise to the Kobayashi-Maskawa, `t Hooft (KMT) effective six-quark interaction, plays a crucial role in determining both the properties of the light mesons and the phase structure of quark matter.  In the first case, the anomaly is responsible for breaking the U(1)$_A$ symmetry of QCD and giving rise to the anomalously large mass of the $\eta^\prime$ meson.  In the second, recent work by Hatsuda \textit{et al.} has demonstrated the importance of the axial anomaly in determining the topology of the QCD phase diagram~\cite{Hatsuda1,Yamamoto}.  In particular, under the right circumstances the anomaly-induced attraction between chiral and diquark condensates leads to a low temperature ($T$) critical point and a corresponding BCS-BEC crossover between the chirally broken Nambu-Goldstone (NG) phase and a color superconducting (CSC) phase characterized by diquark pairing, as shown in Fig.~\ref{fig:schematic}.

One useful method for describing dense quark matter is the Nambu--Jona-Lasinio (NJL) model~\cite{Abuki_AA,Hatsuda2,Buballa,Schafer2,Christov,Warringa,Boomsma2}.  First developed to describe chiral symmetry breaking, it has been extended to include diquark pairing and confinement, via the Polyakov loop~\cite{Polyakov,Ratti,Sasaki,Fukushima,Weise,Meisinger,Abuki,Megias}.  The resulting PNJL model has been shown to possess coincident chiral and deconfinement transitions for both two- and three-flavor systems, in close agreement with current lattice simulations~\cite{Banks,Kogut,Nakano,Bazavov}.  More recently, the NJL model has also been used to investigate the effects of the axial anomaly on the QCD phase diagram~\cite{Abuki_AA}.

In the present paper we further develop recent extensions of the NJL model by including the effects of both the axial anomaly and confinement on the massless three-flavor QCD phase diagram.  In Sec.~\ref{sec:NJLAA} we review the NJL model including the KMT interaction, while in Sec.~\ref{sec:Ploop} we introduce the Polyakov loop and discuss its ability to describe confinement.  In Sec.~\ref{sec:PNJLAA} we combine these two models and derive the thermodynamic potential of the three-flavor PNJL model.  In Sec.~\ref{sec:results} we construct the QCD phase diagram by minimizing the thermodynamic potential, paying special attention to the recently discovered low $T$ critical point and the criteria for its emergence.  Finally, in Sec.~\ref{sec:LG} we perform an expansion of the thermodynamic potential of the Ginzburg-Landau form and assess the lowest-order couplings between the quark condensates and the Polyakov loop.

\section{Three-flavor NJL Model with Axial Anomaly \label{sec:NJLAA}}

Following Abuki \textit{et al.} we write the three-flavor NJL Lagrangian~\cite{Abuki_AA}
\beq
\Lc_{NJL} = \overline{q} (i \slashed{\partial} - \hat{m} + \mu \gamma^0) q + \Lc^{(4)} + \Lc^{(6)} ,
\eeq
where $q = (\mbox{u,d,s})^T$ is the quark field, $\hat{m} = \mbox{diag} (m_u,m_d,m_s)$ is the quark mass matrix in flavor space, which we choose to be diagonal ($\hat{m} = m \mathbf{I}$), $\mu$ is the quark chemical potential, and $\Lc^{(4)}$ and $\Lc^{(6)}$ are four- and six-quark interaction terms respectively.  

The standard form of the four-quark interaction $\Lc^{(4)}$ is chosen to respect the chiral symmetry of QCD and contains color-flavor-locked (CFL) quark-quark interactions~\cite{Pisarski,Schafer}
\beqa
\Lc^{(4)} & = & \Lc^{(4)}_\sigma + \Lc^{(4)}_d   ,   \\
\Lc^{(4)}_\sigma & = & G \sum^8_{a=0} [ (\overline{q} \tau_a q)^2 + (\overline{q} i \gamma_5 \tau_a q)^2   ,   \\
\Lc^{(4)}_d & = & H \sum_{A,A^\prime=2,5,7} [ (\overline{q} i \gamma_5 \tau_A \lambda_{A^\prime} C \overline{q}^T) (q^T C i \gamma_5 \tau_A \lambda_{A^\prime} q)   \nonumber \\
	&& \hspace{20mm} + (\overline{q} \tau_A \lambda_{A^\prime} C \overline{q}^T) (q^T C \tau_A \lambda_{A^\prime} q) ]   ,
\eeqa
where $\tau_a$ are the U(3) flavor generators ($a = 0...8$), $\tau_A$ and $\lambda_A$ are the antisymmetric flavor and SU(3) color generators ($A,A^\prime = 2,5,7$), and $C$ is the charge conjugation operator.  The generators are normalized by the condition $\mbox{Tr} (\tau_a \tau_b) = 2 \delta_{ab}$, and we take $G,H > 0$, corresponding to attractive interactions.  We also find it useful to define the chiral and diquark operators
\beqa
\phi_{ij} & = & (\overline{q}_R)^j_a (q_L)^i_a   ,   \\
(d_L)_{ai} & = & \epsilon_{abc} \epsilon_{ijk} (q_L)^j_b C (q_L)^k_c   ,   \\
(d_R)_{ai} & = & \epsilon_{abc} \epsilon_{ijk} (q_R)^j_b C (q_R)^k_c   ,
\eeqa
where $i,j,k$ and $a,b,c$ are flavor and color indices respectively and $q_L$ and $q_R$ denote states of left- and right-handed chirality.  In terms of these operators, the four-quark interactions can be written in the form
\beqa
\Lc^{(4)}_\sigma & = & 8 G \mbox{Tr}_f (\phi^\dagger \phi)   ,   \\
\Lc^{(4)}_d & = & 2 H \mbox{Tr}_f (d^\dagger_L d_L + d^\dagger_R d_R) .
\eeqa

The six-quark interaction can also be written as the sum of two terms
\beqa
\Lc^{(6)} & = & \Lc^{(6)}_\sigma + \Lc^{(6)}_{\sigma d}   ,   \\
\Lc^{(6)}_\sigma & = & -8 K (\det \phi + \mbox{h.c.})   ,   \\
\Lc^{(6)}_{\sigma d} & = & K^\prime \left( \mbox{Tr}_{c,f} [(d^\dagger_R d_L) \phi] + \mbox{h.c.} \right)   .
\eeqa
The first term, $\Lc^{(6)}_\sigma$, is the standard KMT interaction, which is the result of the instanton-induced QCD axial anomaly~\cite{Kobayashi,tHooft}.  The second term, $\Lc^{(6)}_{\sigma d}$, is the effective interaction between the chiral and diquark fields, mediated by the QCD instanton~\cite{Hatsuda1,Yamamoto}.

The condensates favored by $\Lc^{(4)}$ and $\Lc^{(6)}$ are the flavor-symmetric chiral and diquark condensates in the spin-parity $0^+$ channel:
\beqa
\braket{\overline{q}^i_a q^j_a} & = & \sigma \delta_{ij}   \hspace{10mm} \braket{q^T C \gamma_5 \tau_A \lambda_{A^\prime} q} = d \delta_{A A^\prime}   .
\eeqa
In mean field the Lagrangian becomes
\beqa
\Lc_{NJL} & = & \overline{q} (i \slashed{\partial} - M + \mu \gamma^0 ) q - \frac{1}{2} ( \Delta^\ast q^T C \gamma_5 \tau_A \lambda_A q + \mbox{h.c.} ) \nonumber \\
    && \hspace{5mm} - V(\sigma,d)   ,   \label{eq:Lagrangian}
\eeqa
where the sum over $A = 2,5,7$ is implied and where the effective quark mass is
\beq
M = m - 4 G \sigma + 2 K \sigma^2 + \frac{K^\prime}{4} \hspace{.5mm} |d|^2   ,
\eeq
and the pairing gap $\Delta$ is
\beq
\Delta = - 2 d \left(H - \frac{K^\prime}{4} \hspace{.5mm} \sigma \right)   .
\eeq
In addition, the condensates contribute to the system's potential directly an amount
\beq
V(\sigma,d) = 6 G \sigma^2 + 3 H |d|^2 - 4 K \sigma^3 - \frac{3 K^\prime}{2} \hspace{.5mm} |d|^2 \sigma   .   \label{eq:V(sigma,d)}
\eeq
In order to diagonalize the Lagrangian, it is convenient to define the Nambu-Gor'kov field
\beq
\Psi = \frac{1}{\sqrt{2}} \left( \begin{array}{c} q \\ q^C \end{array} \right)   ,
\eeq
where $q^C = C \overline{q}^T$ is the charge-conjugate quark field.  With this, we write the mean field Lagrangian in the form $\Lc_{NJL} = \overline{\Psi} S^{-1} \Psi - V$, where the inverse propagator in Nambu-Gor'kov space is
\beq
S^{-1} (k) = \left(\begin{array}{cc} \slashed{k} + \mu \gamma^0 - M & \Delta \gamma_5 \tau_A \lambda_A \\
			- \Delta^\ast \gamma_5 \tau_A \lambda_A & \slashed{k} - \mu \gamma^0 - M \end{array} \right)   .   \label{eq:NJLprop}
\eeq
Having diagonalized the Lagrangian, we compute the thermodynamic potential by performing the Gaussian integrals over the fields $\overline{\Psi}$ and $\Psi$.  Thus, we write the thermodynamic potential in the standard way
\beqa
\Omega_{NJL} & = & V(\sigma,d) - \frac{1}{\beta} \sum_n \int \frac{d^3 \mathbf{k}}{(2\pi)^3} \hspace{.5mm} \frac{1}{2} \hspace{.5mm} \tr \ln [\beta S^{-1} (\omega_n,\mathbf{k})]   ,   \nonumber \\ \label{eq:Omega}
\eeqa
where $\beta  = 1/T$ is the inverse temperature, the trace is taken over Dirac, color, flavor, and Nambu-Gor'kov indices, the factor of 1/2 accounts for the double-counting inherent in the Nambu-Gor'kov formalism, and $\omega_n = i (2n+1) \pi T$ are the fermionic Matsubara frequencies.

The couplings $G$, $H$, and $K$, as well as the cutoff $\Lambda$ are fixed by fitting to experimentally determined mesonic properties, as discussed by Buballa~\cite{Buballa}, and are given in Table \ref{tab:couplings}.  The second anomaly coupling $K^\prime$ will be adjusted ``by hand" to determine its effect on the low $T$ critical point.  In the following calculations we assume that the current masses of the quarks are zero.

\begin{center}
\begin{table}
\caption{\footnotesize{Coupling constants and dynamical quark mass for the PNJL model~\cite{Buballa,Abuki_AA}}.}
\vspace{-3mm}
\begin{tabular}{p{20mm}p{20mm}p{20mm}p{20mm}}
\hline \hline
\hspace{1mm}$G \Lambda^2$ \hspace{2mm}  &  \hspace{2mm}$H \Lambda^2$  &  \hspace{7mm}$K \Lambda^5$ \hspace{2mm} & \hspace{5.5mm} $M$ (MeV)   \\
\hline
1.926  &  \hspace{1.5mm}1.74 &  \hspace{6mm}12.36 &  \hspace{9mm}355.1   \\
\hline \hline
\end{tabular}
\label{tab:couplings}
\end{table}
\end{center}

\section{Confinement and the Polyakov loop \label{sec:Ploop}}

Having described the portion of the Lagrangian responsible for chiral symmetry breaking and diquark pairing, we next extend our model to include confinement.  We begin by introducing a temporal gauge field $A_0$, and replacing the derivatives in Eq. (\ref{eq:Lagrangian}) with the covariant form
\beq
D_\mu = \partial_\mu - i A_\mu   \hspace{10mm}   A_\mu = \delta^0_\mu A_0   ,
\eeq
where $A_0 (\mathbf{x},t) \equiv A_0$, a matrix in color space, is chosen to be a time-independent homogeneous gauge field.

As shown by Polyakov, in the absence of quarks the order parameter for confinement is the thermal Wilson line with periodic boundary conditions (also called the Polyakov loop)~\cite{Polyakov,Megias,Svetitsky}.  Unfortunately, despite the ubiquity of this object, its notation in the literature is very inconsistent.  Thus, we define precisely the quantity we will use to describe the QCD deconfinement transition.

The gauge-invariant Wilson loop is defined as
\beq
L(\mathbf{x}) = {\cal{P}} \hspace{.5mm} \mbox{exp} \left \{ i \oint d x^\mu \hspace{.5mm}  A_\mu \right \}    ,
\eeq
where ${\cal{P}}$ is the path-ordering operator.  Note that $L(\mathbf{x})$ inherits any non-Dirac indices of the gauge field $A_\mu$.  Thus, in the present theory, $L(\mathbf{x})$ is a matrix in color space.  When working at finite temperature, we perform a Wick rotation to Euclidean time by defining $\tau = i t$, with the $A_0$ component of the gauge field also rotated to $A_4 = i A_0$.  In the absence of the spatial components of $A_\mu$, the Polyakov-loop matrix is now
\beq
L (\mathbf{x}) = {\cal{P}} \hspace{.5mm} \mbox{exp} \left \{ i \int^\beta_0 d \tau \hspace{.5mm} A_0 (\mathbf{x},\tau) \right \}   ,   \label{eq:loop}
\eeq
where ${\cal{P}}$ is now understood to be the ordering operator in imaginary time.  Specializing to constant $A_0$, Eq.~(~\ref{eq:loop}) reduces to
\beq
L = e^{i \beta A_0}  .
\eeq
This is the expression that we will use throughout our analysis.  Note that $A_0$ is a Hermitian field, in contrast to some authors who write $L = e^{\beta A_0}$, with $A_0$ anti-Hermitian~\cite{Gross,Yaffe}.

In the limit of infinitely massive quarks, the free energy of a static quark-quark pair separated by $\mathbf{x}$ is~\cite{Polyakov,Yaffe}
\beq
e^{-\beta F_{qq} (\mathbf{x})} = \braket{\overline{\Phi}(\mathbf{x}) \Phi(\mathbf{0})}   ,
\eeq
where $\Phi$ and $\overline{\Phi}$ are the thermal expectation values of the traced Polyakov loop and its conjugate
\beq
\Phi (\mathbf{x}) = \frac{1}{N_c} \hspace{.5mm} \braket{ \tr_c L (\mathbf{x})}   \hspace{5mm}   \overline{\Phi} (\mathbf{x}) = \frac{1}{N_c} \hspace{.5mm} \braket{ \tr_c L^\dagger (\mathbf{x})}   .   \label{eq:Ploop}
\eeq
As $|\mathbf{x}| \to \infty$, the expectation values of the operators can be factored to obtain
\beq
e^{-\beta F_\infty} = \braket{\overline{\Phi}(\mathbf{x})} \braket{\Phi(\mathbf{0})} = |\Phi (\mathbf{0})|^2   .
\eeq
For confined quarks, as the separation between two quarks tends to infinity, so does the energy required to maintain the separation ($F_\infty = \infty$).  Thus, $\Phi$ is an order parameter for the deconfinement transition:
\beqa
\Phi & = & 0   \hspace{5mm}   \mbox{confined}   ,   \nonumber   \\
\Phi & \neq & 0   \hspace{5mm}   \mbox{deconfined}   .   \nonumber	
\eeqa
From the normalization adopted in Eq. (\ref{eq:Ploop}) we see that ``complete deconfinement" is the limit $\Phi \to 1$, while $0 < \Phi < 1$ describes a system somewhere between complete confinement and complete deconfinement.
\vspace{20mm}
\begin{center}
\begin{table}[b]
\caption{\footnotesize{Coefficients of the Polyakov-loop potential~\cite{Weise}.}}
\begin{tabular}{p{20mm}p{20mm}p{20mm}p{20mm}}
\hline \hline
$\hspace{1mm}a_0$ & \hspace{3mm}$a_1$ & \hspace{7mm}$a_2$ & \hspace{15mm}$b_3$    \\
\hline
3.51 &  -2.47 &  \hspace{5.5mm}15.2 &  \hspace{12mm}-1.75  \\
\hline \hline
\end{tabular}
\label{tab:coefficients}
\end{table}
\end{center}
\vspace{-29mm}

\begin{figure}
\includegraphics[scale=0.7]{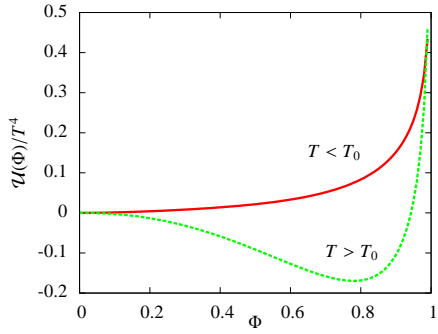}
\caption{\footnotesize{(color online) Dimensionless Polyakov loop potential, ${\cal{U}} (\Phi) / T^4$, which models confinement in the pure-gauge sector. Below $T_0$ the potential is minimized for $\Phi = 0$ and the system is confined, spontaneously breaking QCD's Z(3) center symmetry.  Above $T_0$ the symmetry is restored and the system has a non-zero $\Phi$, indicating movement toward a quark-gluon plasma (QGP).}}
\label{fig:Uplot}
\end{figure}

We next introduce a Polyakov-loop potential $\Uc (\Phi,\overline{\Phi})$ which describes the deconfinement transition in the pure-gauge sector.  Different forms have been proposed for $\Uc (\Phi,\overline{\Phi})$, but in all cases, the functional form is chosen to be consistent with the Z(3) center symmetry of QCD~\cite{Sasaki,Weise,Ratti}.  In this study we follow Fukishima and R\"{o}{\ss}ner and use the potential~\cite{Fukushima,Weise}
\beqa
\frac{\Uc (\Phi,\overline{\Phi})}{T^4} & = & - \frac{1}{2} \hspace{.5mm} a(T) \overline{\Phi} \Phi + b (T) \hspace{.5mm} \ln [ 1 - 6 \overline{\Phi} \Phi \nonumber \\
	&& \hspace{15mm} + 4 (\Phi^3 + \overline{\Phi}^3) - 3 (\overline{\Phi} \Phi)^2 ]  ,   \label{eq:PloopPot}
\eeqa
where the temperature-dependent coefficients have the form
\beq
a (T) = a_0 + a_1 \left(\frac{T_0}{T} \right) + a_2 \left(\frac{T_0}{T} \right)^2   \hspace{5mm}   b (T) =  b_3 \left(\frac{T_0}{T} \right)^3   ;   \nonumber \\
\eeq
the $a_i$ and $b_i$ are fixed by comparison with lattice data at $\mu=0$, and are shown in Table \ref{tab:coefficients}~\cite{Weise}.  The parameter $T_0$, which is 270 MeV in the pure-gauge sector, will be fixed to reproduce the three-flavor lattice transition temperature~\cite{Fukushima2,Karsch,Aoki1,Aoki2}.  Note that, as shown in Fig.~\ref{fig:Uplot}, by diverging in the limit $\Phi \to 1$, the logarithmic form of ${\cal{U}} (\Phi,\overline{\Phi})$ ensures that $\Phi < 1$.

\section{The PNJL Model : Confinement and Chiral Symmetry \label{sec:PNJLAA}}

Having modeled both chiral symmetry breaking and confinement, we now combine the results of the prior sections to write the full Lagrangian $\Lc_{PNJL} = \Lc_{NJL} - \Uc (\Phi,\overline{\Phi})$, with the replacement $\partial_\mu \to D_\mu = \partial_\mu - i A_\mu$:
\beq
\Lc_{PNJL} = \overline{q} (i \slashed{D} - \hat{m} + \mu \gamma^0) q + \Lc^{(4)} + \Lc^{(6)} - \Uc (\Phi,\overline{\Phi})   .   \nonumber \\   \label{eq:PNJLlagrangian}
\eeq
Note that the coupling between the quarks and the Polyakov loop is determined solely by the covariant derivative, which effectively makes the replacement $\mu \to \mu + A_0$.  Thus, in mean field the inverse propagator in Nambu-Gor'kov space becomes (generalizing Eq. (\ref{eq:NJLprop}))
\beq
S^{-1} (k) = \left(\begin{array}{cc} \slashed{k} + (\mu + A_0) \gamma^0 - M & \Delta \gamma_5 \tau_A \lambda_A \\
			- \Delta^\ast \gamma_5 \tau_A \lambda_A & \slashed{k} - (\mu + A_0) \gamma^0 - M \end{array} \right)   .   \nonumber \\   \label{eq:PNJLprop}
\eeq
As noted, $A_0$ is a matrix in color space, and may therefore be expressed in the form $A_0 = A^a_0 \lambda_a /2$.  In the Polyakov gauge, with $A_0$ diagonal, we write~\cite{Weise,Marhauser}
\beq
A_0 = \phi_3 \lambda_3 + \phi_8 \lambda_8   ,   \label{eq:A0}
\eeq
where $\lambda_3$ and $\lambda_8$ are the symmetric generators of SU(3) color.  In analogy with Eq. (\ref{eq:Omega}), the thermodynamic potential is then
\beqa
\Omega_{PNJL} & = & V(\sigma,d) + \Uc (\phi_3,\phi_8) \nonumber \\
       && \hspace{5mm} - \frac{1}{\beta} \sum_n \int \frac{d^3 \mathbf{k}}{(2\pi)^3} \hspace{.5mm} \frac{1}{2} \hspace{.5mm} \tr \ln [\beta S^{-1} (\omega_n,\mathbf{k})]_{A_0 \to i A_0}    \nonumber \\
\eeqa
where we have gone to the Euclidean signature by taking $A_0 \to i A_0$, as is required at finite temperature~\cite{Kapusta,Gjestland}, and have expressed $\Uc (\Phi,\overline{\Phi})$ in terms of the parameters $\phi_3$ and $\phi_8$, which are related via the transformation
\beq
\Phi = \frac{1}{3} \left[ e^{i \beta (\phi_3 + \phi_8 / \sqrt{3})} + e^{i \beta (-\phi_3 + \phi_8 / \sqrt{3})} + e^{-2 i \beta \phi_8 / \sqrt{3}} \right]   .   \nonumber \\   \label{eq:Phiphi}
\eeq

From Eqs. (\ref{eq:PNJLprop}) and (\ref{eq:A0}) we find that the presence of $A_0$ effectively renders the chemical potential color-dependent.  In the Gell-Mann basis, in color space we can write $\mu + i A_0 = \mbox{diag} (\mu_1, \mu_2, \mu_3)$, where
\beqa
\mu_1 & \equiv & \mu + i \left(\phi_3 + \frac{\phi_8}{\sqrt{3}} \right)   ,   \nonumber \\
\mu_2 & \equiv & \mu + i \left(- \phi_3 + \frac{\phi_8}{\sqrt{3}} \right)   ,   \\
\mu_3 & \equiv & \mu - \frac{2 i \phi_8}{\sqrt{3}}   .   \nonumber
\eeqa
The inverse propagator in Eq. (\ref{eq:PNJLprop}) is a $72 \times 72$ matrix, and the determinant is therefore a 72nd-order polynomial in $\omega_n$.  While prior work has investigated special cases of this expression, including the two-flavor PNJL model~\cite{Weise} and the three-flavor NJL model~\cite{Abuki_AA}, in this study we consider the general three-flavor PNJL model.  As a result, the evaluation of the $\mbox{tr} \ln$ in Eq. (\ref{eq:Omega}) is much more difficult, and relies on general relations involving the determinants of $2 \times 2$ (Nambu-Gor'kov and Dirac) and $3 \times 3$ (color and flavor) block matrices.

Exchanging the order of the eigenvalue and Matsubara sums, we make use of the relation
\beq
\sum_n \ln [\beta (\omega_n \pm E_j)] = \ln (1 + e^{-\beta E_j}) + \frac{1}{2} \hspace{.5mm} \Delta E_j   ,
\eeq
where $\Delta E_j = E_j - E^{free}_j$, the difference in the eigenvalue from the noninteracting case, comes from measuring $\Omega$ relative to the noninteracting Dirac sea.  Thus, we are able to write the thermodynamic potential in the form
\beqa
\Omega_{PNJL} & = & V(\sigma,d) + \Uc (\phi_3,\phi_8) \nonumber \\
	&& - \frac{1}{2 \beta} \sum_j \int \frac{d^3 \mathbf{k}}{(2\pi)^3} \left[ \ln (1 + e^{-\beta E_j}) + \frac{1}{2} \hspace{.5mm} \beta \Delta E_j \right]   .   \nonumber \\
	\label{eq:PNJLomega}
\eeqa
Solving the characteristic polynomial of Eq. (\ref{eq:PNJLprop}) yields 18 distinct eigenvalues, each with multiplicity 4 (2 spin $\times$ 2 Nambu-Gor'kov).  The first 48 eigenvalues (of which 12 are distinct) are of the form
\beq
E_{1-48} = \sqrt{ \left(E_k \pm \frac{\mu_i + \mu_j}{2} \right)^2 + |\Delta|^2} \pm \frac{\mu_i - \mu_j}{2}   ,   \label{eq:eigenvals1}
\eeq
where $i \neq j$, $E_k = \sqrt{\mathbf{k}^2 + M^2}$, and the two $\pm$ are independent.  The remaining 24 eigenvalues (6 distinct) are the roots of the polynomials $F(\omega_n,E_k)$ and $F(\omega_n,-E_k)$ which satisfy $\lim_{k \to \infty} E_j = \infty$, where
\begin{widetext}
\beqa
F(\omega_n,E_k) & = & (\omega_n + E_k + \mu_1) (\omega_n - E_k - \mu_1) (\omega_n + E_k + \mu_2) (\omega_n - E_k - \mu_2) (\omega_n + E_k + \mu_3) (\omega_n - E_k - \mu_3)   \nonumber \\
	&& \hspace{5mm} - |\Delta|^2 (\omega_n + E_k + \mu_1) (\omega_n + E_k + \mu_2) (\omega_n - E_k - \mu_3) [ (\omega_n - E_k - \mu_1) + (\omega_n - E_k - \mu_2) ]   \nonumber \\
	&& \hspace{5mm} - |\Delta|^2 (\omega_n + E_k + \mu_1) (\omega_n + E_k + \mu_3) (\omega_n - E_k - \mu_2) [ (\omega_n - E_k - \mu_1) + (\omega_n - E_k - \mu_3) ]   \nonumber \\
	&& \hspace{5mm} - |\Delta|^2 (\omega_n + E_k + \mu_2) (\omega_n + E_k + \mu_3) (\omega_n - E_k - \mu_1) [ (\omega_n - E_k - \mu_2) + (\omega_n - E_k - \mu_3) ]   \nonumber \\
	&& \hspace{5mm} + |\Delta|^4 [ (\omega_n + E_k + \mu_1) + (\omega_n + E_k + \mu_2) + (\omega_n + E_k + \mu_3) ]   \nonumber \\
	&& \hspace{15mm} \times [ (\omega_n - E_k - \mu_1) + (\omega_n - E_k - \mu_2) + (\omega_n - E_k - \mu_3) ] - 4 |\Delta|^6   .   \label{eq:eigenvals2}
\eeqa
\end{widetext}
The function $F(\omega_n,E_k)$ is an even sixth-order polynomial and therefore reducible to a cubic polynomial, whose solutions can be obtained exactly.  Thus, all of the model's eigenvalues can be obtained explicitly, and the thermodynamic potential computed via Eq. (\ref{eq:PNJLomega}).  In order to construct the phase diagram, we must now minimize $\Omega (\sigma,d,\phi_3,\phi_8)$ with respect to all variables.

We must first ensure that the minimization of $\Omega$ is well-defined.  In fact, from Eqs. (\ref{eq:eigenvals1}) and (\ref{eq:eigenvals2}) we see that the eigenvalues are manifestly complex, and the imaginary part of $\Omega$ need not vanish.  In general, $\Omega$ is therefore a complex function, the minimization of which is ill-defined. Indeed, it is not at all clear how one would interpret a complex potential in this context.  However, as noted by Weise \textit{et al}. in the context of the two-flavor PNJL model, if we assume that $\phi_3$ and $\phi_8$ are real then it follows that $\phi_8 = 0$, so that only $\phi_3$ is left as an independent variable~\cite{Weise}.  Then Eq. (\ref{eq:Phiphi}) reduces to
\beq
\Phi = \frac{1 + 2 \cos (\beta \phi_3)}{3}   .   \label{eq:phiPhi}
\eeq
For $\phi_8 = 0$ we also find $\mu^\dagger_1 = \mu_2$ and $\mu_3 = \mu$, so the eigenvalues come in conjugate pairs. This ensures that the thermodynamic potential is in fact real and that the minimization procedure is well-defined.

\section{Phase Structure \label{sec:results}}

Having obtained the thermodynamic potential, we now construct the 3-flavor PNJL model phase diagram by minimizing Eq. (\ref{eq:PNJLomega}) with respect to $\sigma$, $d$, and $\phi_3$ at each point ($\mu,T$).  In order to assess the effects of confinement on the QCD phase structure and the low $T$ critical point we first construct the phase diagram for $\Phi = 1$ (no confinement), and then compare to the results for the full PNJL model.

\subsection{Without Confinement}

\begin{figure}
\includegraphics[scale=0.8]{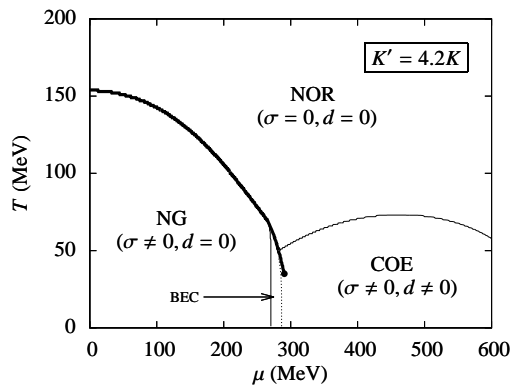}
\caption{\footnotesize{Phase diagram for the three-flavor NJL model (no confinement).  Thick lines represent first-order phase transitions while thin lines represent second order transitions.  The dotted vertical line is BEC-BCS crossover defined by $M (\mu,T) = \mu$.  The low $T$ critical point is at $(\mu,T) = (291 \mbox{ MeV, } 35 \mbox{ MeV})$.}}
\label{fig:njlPD}
\end{figure}

In the absence of the Polyakov loop $A_0 = \phi_3 = 0$, so that we set $\Phi = 1$ (see Eq. (\ref{eq:phiPhi})) and eliminate ${\cal{U}} (\Phi,\overline{\Phi})$ from the thermodynamic potential.  The thermodynamic potential then reduces to
\beqa
\Omega_{NJL} & = & V(\sigma,d) - 2 T \sum_\pm \int \frac{d^3 \mathbf{k}}{(2\pi)^3} \bigg[ 8 \ln (1 + e^{-\beta E^\pm_1})   \nonumber \\
	     &&   \hspace{10mm} + \ln (1 + e^{-\beta E^\pm_2}) + 4 \beta \Delta E^\pm_1 + \frac{1}{2} \hspace{.5mm} \beta \Delta E^\pm_2 \bigg]   ,   \nonumber \\
\eeqa
where the eigenvalues are now
\beqa
E^\pm_1 & = & \sqrt{(E_k \pm \mu)^2 + |\Delta|^2}   \\
E^\pm_2 & = & \sqrt{(E_k \pm \mu)^2 + 4 |\Delta|^2} ,
\eeqa
and where $\sum_\pm$ denotes summation over $E^{+}_{1,2}$ and $E^{-}_{1,2}$.

\begin{figure}
\includegraphics[scale=0.8]{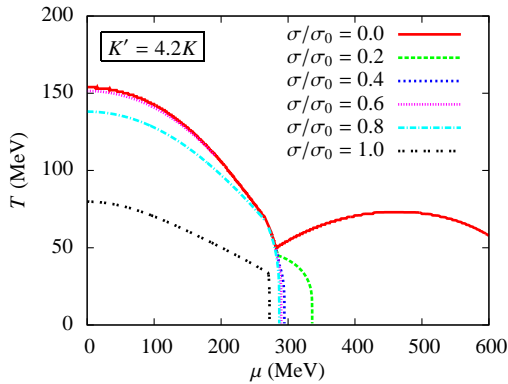}
\caption{\footnotesize{(color online) Contour plot for the chiral condensate in the NJL model, showing contours of $\sigma$ = 0.2, 0.4, 0.6, 0.8 and 1.0 $\sigma_0$, where $\sigma_0$ is the maximum value of the chiral condensate.}}
\label{fig:njl-sigmacontours}
\end{figure}

The minimization of $\Omega$ with respect to $\sigma$ and $d$ yields the phase diagram shown in Fig. \ref{fig:njlPD}.  The critical temperature at $\mu = 0$ is found to be $T^{\tiny{NJL}}_c = 153 \mbox{ MeV}$, in agreement with Abuki \textit{et al.}, and within the margin of error of current lattice results~\cite{Abuki_AA,Aoki1,Aoki2}.  Note that this temperature is determined entirely by the couplings $G$, $H$, and $K$, which were in turn fixed by emperical mesonic properties ($T^{\tiny{NJL}}_c$ proves independent of $K^\prime$, which describes condensate coupling, as the diquark condensate is absent in this portion of the phase diagram).  

As reported by Abuki \textit{et al}., the topology of the phase diagram depends critically on the ratio of anomaly couplings $\kappa = K^\prime / K$~\cite{Abuki_AA}.  We find, in agreement with their report, that for $\kappa < 4.2$, the transition out of the Nambu-Golstone (NG) phase is first-order for all temperatures, while for $\kappa \geq 4.2$, the low-$T$ critical point emerges, below which there is a smooth crossover from the NG phase to a CSC-like COE phase.

In a result not previously reported, we find that as $\kappa$ increases, the critical point moves up the NG-COE phase boundary (to higher $T$ and lower $\mu$) until it vanishes into the NG-``normal" (NOR) phase boundary at $\kappa = 4.8$.  We therefore find that there are three distinct structures of the NJL phase diagram, determined by the value of $\kappa$: (1) $\kappa < 4.2$, (2) $4.2 \leq \kappa \leq 4.7$, and (3) $\kappa \geq 4.8$.  Since we focus on the existence of the low $T$ critical point, we are interested primarily in structure (2).  Therefore, in displaying our results, we choose $\kappa = 4.2$ as a representative value for which the critical point is realized.

\begin{figure}
\includegraphics[scale=0.8]{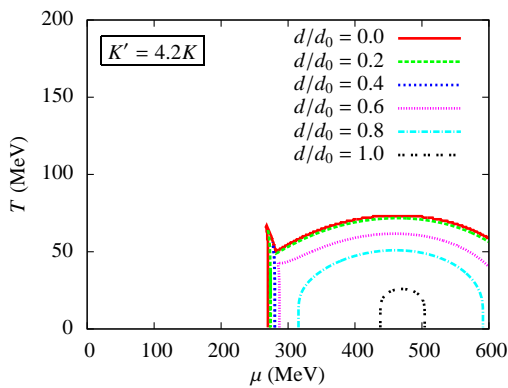}
\caption{\footnotesize{(color online) Contour plot for the diquark condensate in the NJL model, showing contours of $d$ = 0.2, 0.4, 0.6, 0.8 and 1.0 $d_0$, where $d_0$ is the maximum value of the diquark condensate.}}
\label{fig:njl-dcontours}
\end{figure}

Considering the contour plot of $\sigma$, shown in Fig.~\ref{fig:njl-sigmacontours}, we see that the chiral condensate is relatively slowly varying within the NG phase, which corresponds to a relatively constant effective quark mass.  Near the phase boundaries, however, $\sigma$ varies rapidly, either falling discontinuously (NG-NOR or NG-COE, above the critical point) or undergoing a very rapid, though smooth, crossover (NG-COE, below the critical point). 

Similarly, the contours of $d$ show that while $d$ is not maximized over as great a portion of the phase diagram as $\sigma$, it is relatively constant throughout the COE region, dropping rapidly only near the second order COE-NOR phase transition and the NG-COE crossover (Fig.~\ref{fig:njl-dcontours}).

\subsection{With Confinement}

Before constructing the three-flavor QCD phase diagram with confinement, as noted in Sec.~\ref{sec:Ploop}, we fix the parameter $T_0$ by matching the critical temperature at $\mu = 0$ with current lattice data.  With massive quarks, the definition of $T_c$ in this context would be, as noted by Aoki \textit{et al.}, ambiguous, there being at least three standard choices: (1) a maximum of the chiral susceptibility, (2) a maximum of the quark number susceptibility, and (3) a maximum in $d \Phi / d T$~\cite{Aoki1,Aoki2,footnote}.  While these transitions are coincident in the NJL and PNJL models, current lattice calculations with non-zero current quark masses yield slightly different values for the three critical temperatures.  However, for massless quarks, all three transitions are coincident and there is no ambiguity.  Thus, we choose to match the deconfinement transition at $T_c = 154$ MeV~\cite{Fukushima2,Karsch}, which leads us to set $T_0 = 50$ MeV.

\begin{figure}
\includegraphics[scale=0.8]{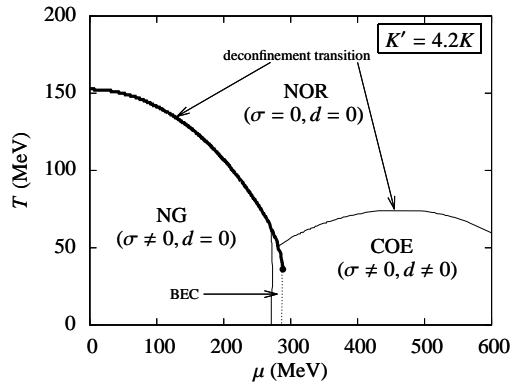}
\caption{\footnotesize{Phase diagram for the three-flavor PNJL model.  Thick lines represent first-order transitions, while thin lines represent second order transitions.  The dotted vertical line is the BEC-BCS crossover, defined by $M(\mu,T) = \mu$.  The low $T$ critical point is at $(\mu,T) = (288 \mbox{ MeV, } 36 \mbox{ MeV}).$}}
\label{fig:phasediagram}
\end{figure}

\begin{figure}
\includegraphics[scale=0.8]{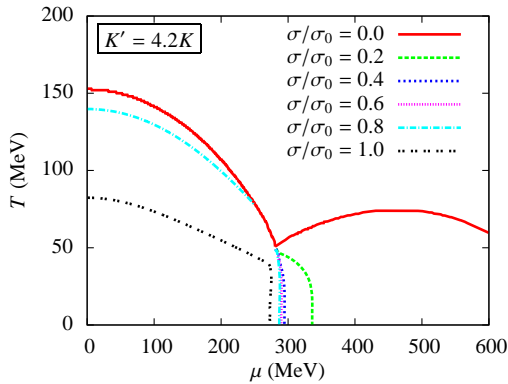}
\caption{\footnotesize{(color online) Contour plot for the chiral condensate in the PNJL model.}}
\label{fig:sigmacontours}
\end{figure}

\begin{figure}
\includegraphics[scale=0.8]{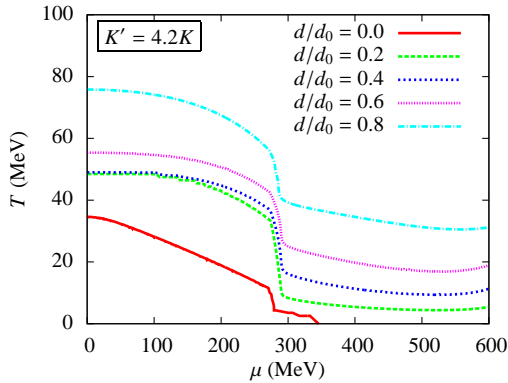}
\caption{\footnotesize{(color online) Contour plot for the traced Polyakov loop in the PNJL model.}}
\label{fig:Phicontours}
\end{figure}

Minimizing $\Omega$ in the presence of the Polyakov loop yields the phase diagram shown in Fig.~\ref{fig:phasediagram}.  Comparing to Fig.~\ref{fig:njlPD} we now assess the effects of confinement on the phase structure of QCD.  As in the nonconfining NJL model, the topology of the phase diagram depends critically on $\kappa$.  We find that this dependence is unaffected by the inclusion of confinement, and that the critical point continues to appear for $4.2 \leq \kappa \leq 4.7$, while it vanishes into the NG-NOR phase boundary for $\kappa \geq 4.8$.  The location in the phase diagram at which the critical point vanishes (for $\kappa = 4.8$) is at marginally higher temperature in the PNJL model ($T = 54$ MeV) than in the NJL model ($T = 50$ MeV).  This reflects the fact that the Polyakov loop causes a slight shift (between 2 and 4 MeV) of the NG-NOR and COE-NOR phase boundaries to higher temperatures, at intermediate to high $\mu$ (note that the critical temperature at $\mu = 0$ is unchanged).

Given that deconfinement is a high $T$ effect, it might be unsurprising that it does not materially affect the low $T$ critical point.   However, it is important to note that were possible $\mu$ dependence included in the Polyakov loop potential, Eq.~(\ref{eq:PloopPot}), the present inclusion of confinement could have a greater effect on the phase structure of QCD.  Unfortunately, because lattice calculations are restricted to $\mu = 0$, we are unable to discern any $\mu$ dependence of $\Uc (\Phi,\overline{\Phi})$.

We also note that while prior work has demonstrated that inclusion of the Polyakov loop pulls the NG-NOR and CSC-NOR phase transitions, as well as the low-$T$ critical point, to higher temperatures~\cite{Weise,Blaschke,Dumm}, our results do not demonstrate such a shift.  This apparent disparity is a result of the aforementioned fitting of $T_0 = 50$ MeV, which we choose to reproduce the $\mu = 0$ deconfinement transition for massless three-flavor QCD.  Had we instead chosen to match the transition in the pure-gauge sector (as in~\cite{Weise}) and set $T_0 = 270$ MeV, both the NG-NOR and CSC-NOR transitions, as well as the low-$T$ critical point, would have been shifted to significantly higher temperatures.

Comparing the contour plot of the chiral condensate (Fig.~\ref{fig:sigmacontours}) to that from the NJL model (Fig.~\ref{fig:njl-sigmacontours}), we note that the Polyakov loop encourages larger values of $\sigma$, most notably near the phase boundaries.  This is clear from the absence of a $\sigma = 0.6 \sigma_0$ contour at low $\mu$ in the PNJL model, as well as the termination of the $\sigma = 0.8 \sigma_0$ contour into the NG-NOR phase boundary at lower chemical potential ($\mu = 250$ MeV) than in the NJL model ($\mu = 266$ MeV).  This effect can be traced to an effective $\sigma^2 \Phi$ coupling, which favors the coexistence of a chiral condensate and confinement, and which is discussed in more detail in Sec.~\ref{sec:LG}.  In the same vein, from Fig.~\ref{fig:Phicontours} we note that $\Phi$ tends to decrease in the presence of $\sigma$.  As a result, curves of constant $\Phi$ are ``pulled" to higher temperatures in the NG phase than they would be in the absence of the effective $\sigma^2 \Phi$ coupling.  Finally, the effective quark mass obtained at ``normal" conditions (i.e., $\mu,T = 0$) is identical ($M = 355$ MeV) whether computed with or without confinement.

\begin{figure}
\includegraphics[scale=0.8]{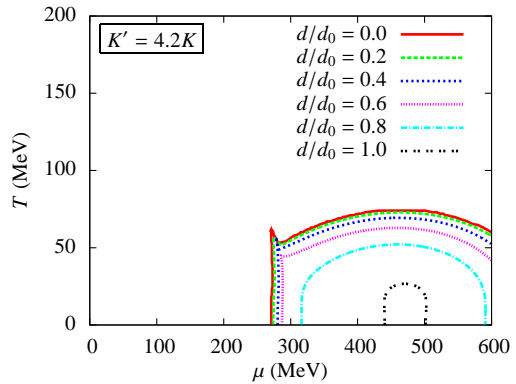}
\caption{\footnotesize{(color online) Contour plot for the diquark condensate in the PNJL model.}}
\label{fig:dcontours}
\end{figure}

\begin{figure}
\includegraphics[scale=0.7]{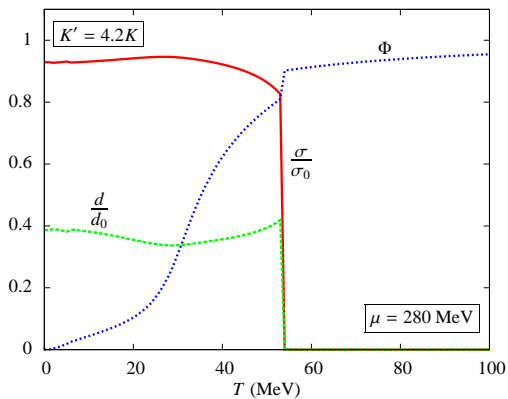}
\caption{\footnotesize{(color online) Plot of the normalized chiral and diquark condensates and Polyakov loop for $\mu = 280$ MeV.}}
\label{fig:mu280}
\end{figure}

Inspecting the contours of the diquark condensate (Fig.~\ref{fig:dcontours}), we find that the Polyakov loop also has no appreciable effect on $d$.

Finally, Fig.~\ref{fig:Phicontours} demonstrates that $\Phi$ is only weakly dependent on chemical potential, being primarily an increasing function of temperature, and what $\mu$ dependence does exist is almost entirely restricted to the NG phase.  In the COE phase, the $\Phi$ contours line up roughly parallel to the $d$ contours, suggesting an effective $|d|^2 \Phi$ coupling.  However, it does not appear that the coupling affects the deconfinement transition significantly, since $\Phi$ has already achieved nearly its maximum value at much lower temperatures than where $d \to 0$.  We find that $\Phi$ is discontinuous across the first-order NG-NOR and NG-COE transitions (e.g. note the jump in the $\Phi = 0$ contour at $\mu = 267$ MeV), but the relative magnitude of the discontinuity is much less than that of $\sigma$ (Fig.~\ref{fig:mu280}).

\section{Ginzburg-Landau Coefficients \label{sec:LG}}

Having constructed the PNJL phase diagram, we next seek to understand the first-order effects of the condensate-Polyakov loop couplings by expanding the thermodynamic potential $\Omega$ in a Ginzburg-Landau (GL) form
\beqa
\Omega & = & \left(\frac{a}{2} \hspace{.5mm} \sigma^2 - \frac{c}{3} \hspace{.5mm} \sigma^3 + \frac{b}{4} \hspace{.5mm} \sigma^4 \right) + \left(\frac{\alpha}{2} \hspace{.5mm} |d|^2 + \frac{\beta}{4} \hspace{.5mm} |d|^4 \right) \nonumber \\ 
       && \hspace{5mm} + \left(A \Phi + \frac{B}{2} \Phi^2 + \frac{C}{3} \hspace{.5mm} \Phi^3 \right) - \gamma \sigma |d|^2   \nonumber \\
       && \hspace{5mm} + \frac{a^\prime}{2} \hspace{.5mm} \sigma^2 \Phi - \frac{c^\prime}{3} \hspace{.5mm} \sigma^3 \Phi + \frac{\alpha^\prime}{2} \hspace{.5mm} |d|^2 \Phi - \gamma^\prime \sigma |d|^2 \Phi + \cdots   \nonumber   \\
	\label{eq:LGomega}
\eeqa
While prior work by Hatsuda \textit{et al.} focused on the topological consequences of a thermodynamic potential of this form (without the Polyakov loop), here we are in a position to compute the coefficients explicitly as functions of temperature and chemical potential~\cite{Hatsuda1}.  For example, the coefficient $a$ can be computed from Eq. (\ref{eq:PNJLomega}) via the relation
\beq
a = \frac{\partial^2 \Omega}{\partial \sigma^2} \bigg|_{\sigma=d=\Phi=0}   ,   \label{eq:coefa}
\eeq
while similar expressions hold for the other coefficients.

In the following calculations, for the sake of convenience the coefficients $a,c,\gamma,$ etc. will be taken to refer to their dimensionless versions, where they are scaled by the appropriate power of $\Lambda$ (as in Table~\ref{tab:couplings}).  This ensures that the coefficients are of roughly the same order of magnitude and facilitates comparison of their relative importance.  For the purposes of these comparisons, note that the dimensionless chiral and diquark condensates have maximum values of $\sigma_0 = 0.0636$ and $d_0 = 0.0548$.  In addition, in order to express the coefficients compactly, we define the following quantities:
\beqa
f^\pm & = & \frac{1}{\cosh(\beta E^\pm_0) - \frac{1}{2}}   ,   \\
g^\pm & = & \frac{1}{\cosh(\beta E^\pm_0) + 1}   ,   \\
h^\pm & = & \frac{1}{9 (\beta E^\pm_0)^2 + \pi^2}   ,
\eeqa
where $E^\pm_0 = k \pm \mu$ are the eigenvalues (without absolute values) in the absence of any interactions.

\subsection{Noncoupling terms}

\begin{figure}
\includegraphics[width=0.40\textwidth,height=0.28\textwidth]{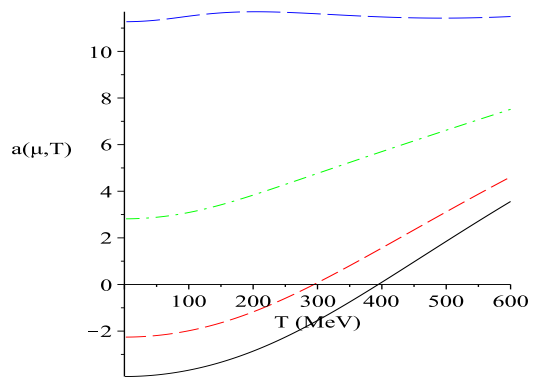}
\caption{\footnotesize{(color online) Coefficient of $\sigma^2$, $a(\mu,T)$, for various values of the chemical potential (solid = 0, dash = 150 MeV, dot-dash = 300 MeV, long-dash = 450 MeV).  For $a < 0$ the $\sigma^2$ term favors spontaneous formation of a chiral condensate.}}
\label{fig:a(T)}
\end{figure}

In computing the coefficients of simple powers of $\sigma$, we may set $d = \Phi = 0$ prior to taking the necessary derivatives.  Thus, the 18 distinct eigenvalues shown in Eqs. (\ref{eq:eigenvals1}) and (\ref{eq:eigenvals2}) reduce to the six distinct values: $E_{1-4} = |E_k \pm \mu| \pm 2 \pi T / 3$ (with the two $\pm$ independent) and $E_{5,6} = |E_k \pm \mu|$.  In this way, we obtain the first three LG coefficients:

\begin{figure*}
\centering
\subfloat{\includegraphics[width=0.45\textwidth,height=0.35\textwidth]{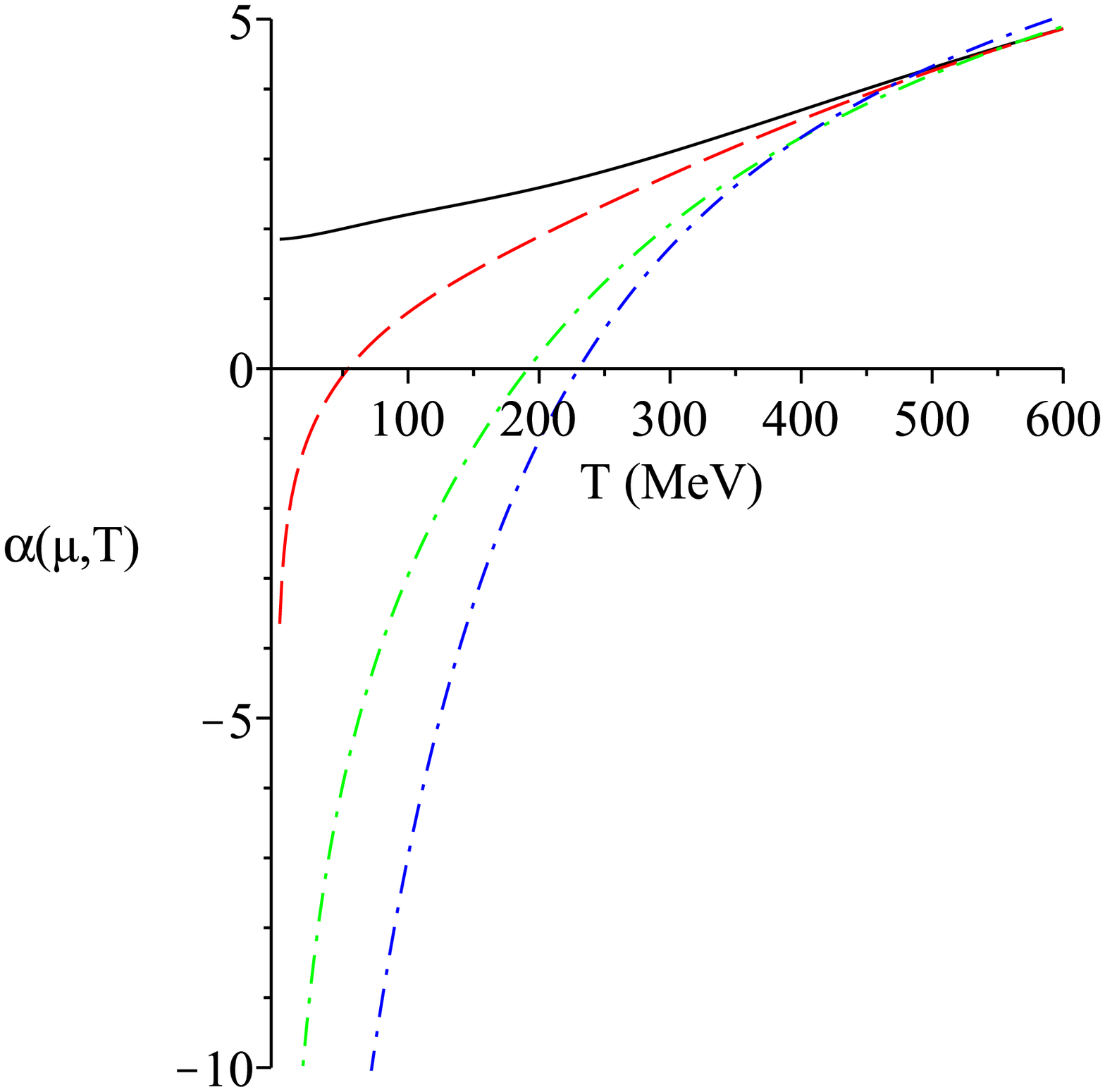} \label{fig:alpha(T)}}
\subfloat{\includegraphics[width=0.45\textwidth,height=0.35\textwidth]{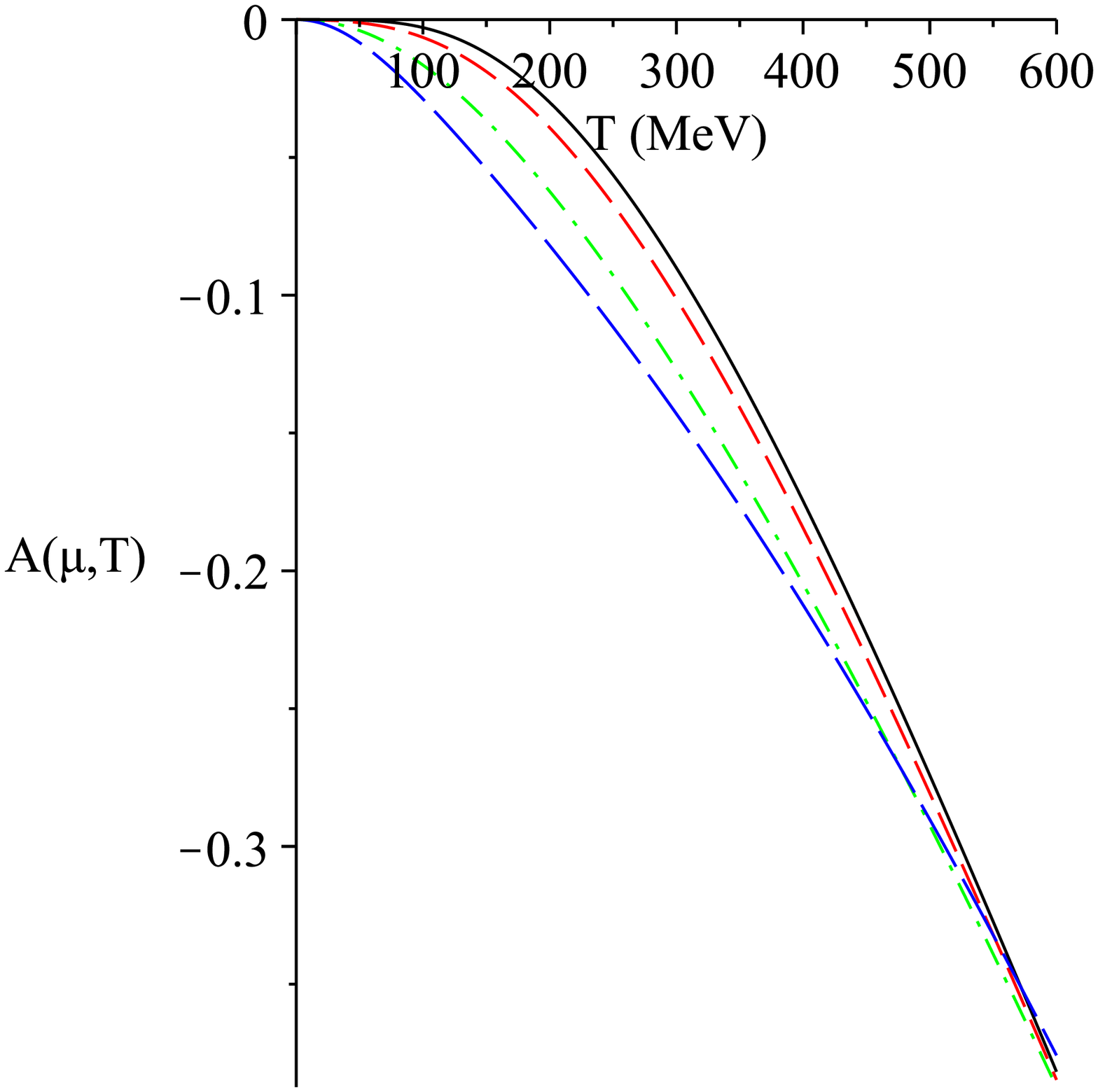} \label{fig:A2(T)}}
\caption{\footnotesize{(color online) (a) Coefficient of $|d|^2$, $\alpha (\mu,T)$, for various values of the chemical potential (same as Fig.~\ref{fig:a(T)}).  For $\alpha < 0$ the $|d|^2$ term favors spontaneous formation of a diquark condensate. (b) Coefficient of $\Phi$, $A (\mu,T)$, for various values of the chemical potential.  For $A(T) < 0$ the linear term favors the formation of a non-zero $\Phi$.}}
\label{fig:aalpha(T)}
\end{figure*}

\beqa
a(\mu,T) & = & 12 G \nonumber \\ && - 48 G^2 \sum_\pm \int \frac{d^3 \mathbf{k}}{(2\pi)^3} \frac{\sinh(\beta E^\pm_0)}{k} ( 2 f^\pm + g^\pm)   ,   \nonumber \\
c (\mu,T) & = & 12 K \nonumber \\ && - 72 G K \sum_\pm \int \frac{d^3 \mathbf{k}}{(2\pi)^3} \frac{\sinh(\beta E^\pm_0)}{k} ( 2 f^\pm + g^\pm) ,   \nonumber \\   \label{eq:coefacb}
\eeqa
\noindent where $\sum_\pm$ denotes summation over $E^{+}_0$ and $E^{-}_0$.  We note that the $\sigma^3$ term is proportional to the coupling $K$, which stems from the axial anomaly.  This proves true for all odd powers of $\sigma$ so that in the absence of the axial anomaly the thermodynamic potential is an even function of $\sigma$.

Figure~\ref{fig:a(T)} shows that the coefficient $a$ changes sign, becoming negative at low temperatures and chemical potentials.  As a result, in the low-$\mu$, low-$T$ portion of the phase diagram, the $\sigma^2$ term in the thermodynamic potential tends to favor chiral condensation.  In fact, we see that if the $\sigma^2$ term were dominant, this transition would occur at $\mu = 0$ at the extremely high temperature of $T = 395 \mbox{ MeV}$.  However, by looking at Eq. (\ref{eq:coefacb}) we can assess the relative magnitude of $c$, and whether it will play a significant role in determining the order of the NG-NOR phase transition.  In fact, noting that the integrals appearing in $c$ are identical to those in $a$, we can write
\beq
c = \frac{3 K}{2 G} (a - 4 G)
\eeq
\hspace{0.1mm}

Thus, we find that at $\mu = 0$, $|c / a| \gtrsim 6 K \sim 60$, while as noted above, $\sigma_0 \sim 1/16$.  As expected from the results of the prior section then, we find that the $\sigma^3$ term cannot be ignored in determining the order of the NG-NOR transition.  In fact, the inclusion of higher-order terms coupling $\sigma$ and $\Phi$ leads to a first-order transition at a more modest temperature of $T_c = 154 \mbox{ MeV}$.

Next, the coefficients of terms involving only powers of $|d|$ may be obtained by setting $\sigma = \Phi = 0$ at the outset and taking the appropriate derivatives.  Unlike the prior calculation, in which setting $d = 0$ reduced the number of distinct eigenvalues from 18 to 6, in this case, no such simplification occurs and the full 18 eigenvalues must be evaluated.  Doing so yields the $|d|^2$ coefficient:

\beqa
\alpha (\mu,T) & = & 6 H - 4 H^2 \sum_\pm \int \frac{d^3 \mathbf{k}}{(2\pi)^3} \bigg \{ \frac{4 \sinh(\beta E^\pm_0) f^\pm}{E^\pm_0} \nonumber \\
	&& + 9 \beta [ 6 \beta E^\pm_0 \sinh(\beta E^\pm_0) + \sqrt{3} \pi] f^\pm h^\pm \nonumber \\
	&& + 36 \beta^2 E^\pm_0 \sinh (\beta E^\pm_0) g^\pm h^\pm \bigg \}  .   \nonumber \\
\eeqa

As shown in Fig.~\ref{fig:alpha(T)}, for sufficiently large $\mu$, the sign of the $|d|^2$ term becomes negative at low $T$.  Since no odd powers of the diquark condensate appear in $\Omega$ the phase transition is of second order.  This transition produces a CSC-like COE phase of quark Cooper pairs in the low $T$, high $\mu$ portion of the phase diagram, as has been widely reported~\cite{Buballa,Schafer2,Weise,Alford}.  The precise location of the diquark condensate formation is affected by the chiral-diquark condensate coupling, but for $T \sim 0$, we find that $\alpha$ becomes negative at $\mu \sim 120 \mbox{ MeV}$.

Next, we move on to compute the coefficients of powers of $\Phi$.  Setting $\sigma = d = 0$ again reduces the 18 eigenvalues to six eigenvalues analogous to those involved in the calculation of the $\sigma^n$ coefficients (with $E_k \to k$).  Computing the necesary derivatives yields:

\begin{figure*}
\centering
\subfloat{\includegraphics[width=0.45\textwidth,height=0.35\textwidth]{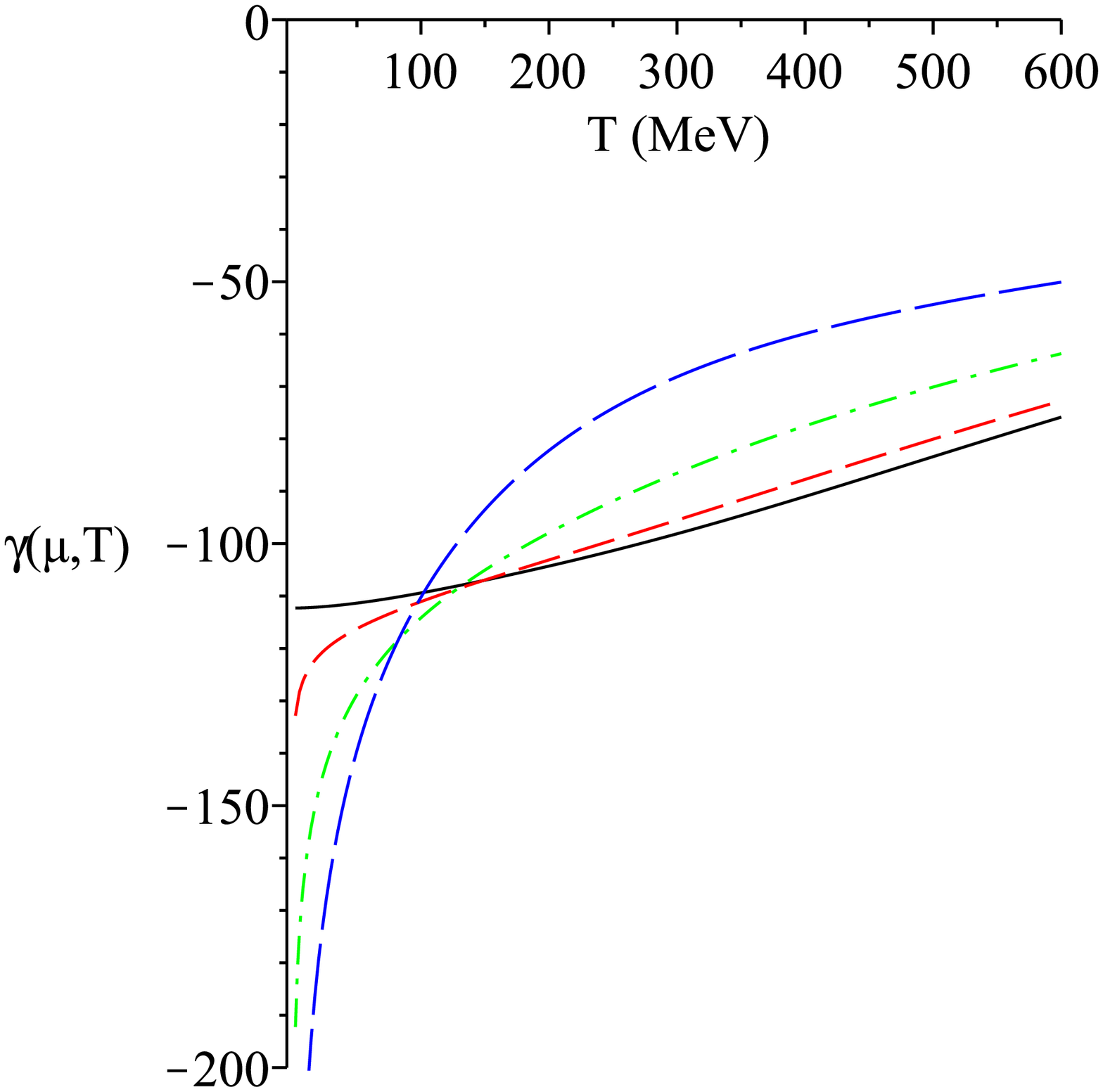} \label{fig:gamma(T)}}
\subfloat{\includegraphics[width=0.45\textwidth,height=0.35\textwidth]{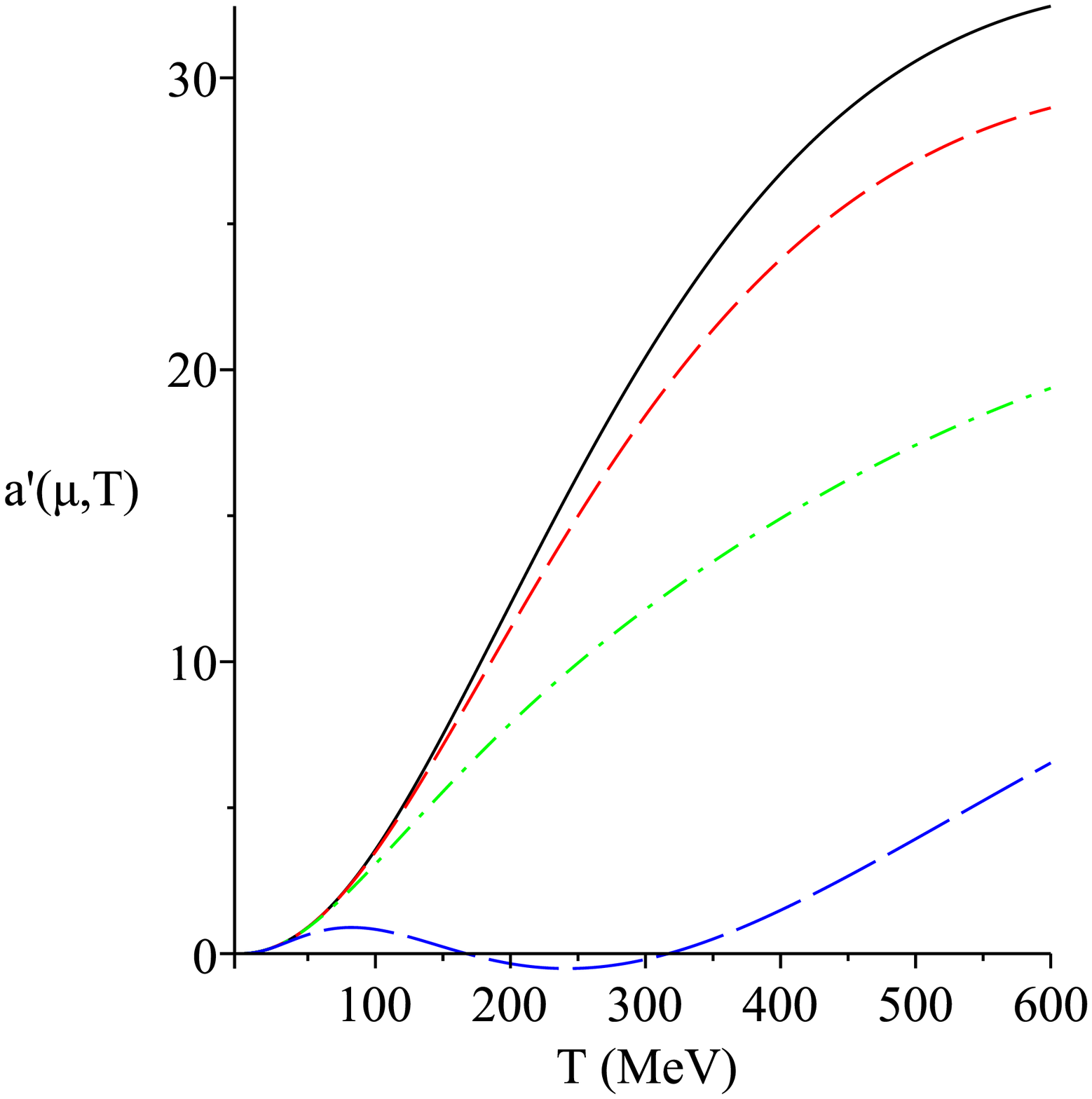} \label{fig:ap(T)}}
\caption{\footnotesize{(color online) (a) The coefficient of $\sigma |d|^2$, $\gamma (\mu,T)$, for various values of the chemical potential (same as Fig.~\ref{fig:a(T)}).  For $\gamma < 0$ the system favors coexistence of chiral and diquark condensates.   (b) Coefficient of $\sigma^2 \Phi$, $a^\prime (\mu,T)$, for various values of the chemical potential.  For $a^\prime < 0$, the $\sigma^2 \Phi$ term favors simultaneous formation of a chiral condensate and deconfinement.}}
\label{fig:Agamma(T)}
\end{figure*}

\beqa
A (\mu,T) & = & - 9 T \sum_\pm \int \frac{d^3 \mathbf{k}}{(2\pi)^3} \hspace{.5mm} f^\pm   ,   \nonumber \\
B (\mu,T) & = & - T^4 \left[ a(T) + 12 b(T) \right] + \frac{27 T}{2} \sum_\pm \int \frac{d^3 \mathbf{k}}{(2\pi)^3} \hspace{.5mm} (f^\pm)^2   ,   \nonumber  \\
C (\mu,T) & = & 24 b(T) - \frac{81 T}{4} \sum_\pm \int \frac{d^3 \mathbf{k}}{(2\pi)^3} \hspace{.5mm} (f^\pm)^3   .   \nonumber  \\
\eeqa

\noindent Significantly, while in the pure-gauge sector the lowest-order term in ${\cal{U}} (\Phi,\overline{\Phi})$ is quadratic (see Eq. (\ref{eq:PloopPot})), the existence of quarks generates a term linear in $\Phi$.  Further, $A$ is negative at all points in the phase diagram except at $T = 0$.  As a result, even for very small non-zero temperatures, the Polyakov loop will take on a finite value.  This behavior stands in marked contrast to that of the pure-gauge sector, in which $\Phi$ undergoes a large discontinuous jump at $T_0 = 270$ MeV, below which $\Phi = 0$.

\subsection{Coupling terms}

Having computed the coefficients of the pure condensate and Polyakov loop terms in Eq. (\ref{eq:LGomega}), we now consider the lowest-order couplings between these variables.  Beginning with the chiral and diquark condensates, we note from Figs.~\ref{fig:sigmacontours} and~\ref{fig:dcontours} that there is only a small region near $\mu \sim 280-320$ MeV in which both $\sigma$ and $d$ are significant.  Thus, the primary condensate coupling is the lowest-order coupling, of the form  $\sigma |d|^2$.  Setting $\Phi = 0$ at the outset and performing the necessary derivatives yields

\begin{widetext}
\beqa
\gamma (\mu,T) & = & \frac{3}{2} \hspace{.5mm} K^\prime - 2 G K^\prime \sum_\pm \int \frac{d^3 \mathbf{k}}{(2\pi)^3} \bigg \{ \frac{3 G \sinh(\beta E^\pm_0)}{k} \hspace{.5mm} (2 f^\pm + g^\pm) + \frac{2 H \sinh (\beta E^\pm_0) f^\pm}{E^\pm_0}  \nonumber \\
	&& \hspace{40mm} + 3 H \beta [ \beta E^\pm_0 \sinh(\beta E^\pm_0) + \sqrt{3} \pi] f^\pm h^\pm  + 18 H \beta^2 E^\pm_0 \sinh (\beta E^\pm_0) h^\pm g^\pm \bigg \}   .   \label{eq:gamma}
\eeqa
\end{widetext}

In Fig.~\ref{fig:gamma(T)} we see that $\gamma$ is negative at all points in the phase diagram, and therefore the $\sigma |d|^2$ coupling universally encourages coexistence of the chiral and diquark condensates.  In the prior section we observed that this term is the critical factor in determining the nature of the NG-COE transition at low temperatures.  Since the $\sigma |d|^2$ coupling does not involve the Polyakov loop, Eq. (\ref{eq:gamma}) is the same as that obtained by Abuki \textit{et al.}, although they did not compute it explicitly, but only observed its consequences in the numerical construction of the QCD phase diagram~\cite{Abuki_AA}.

Having computed the lowest-order noncoupling terms, we consider the effective modifications to these terms that arise from the inclusion of the Polyakov loop.  The coefficients of the lowest-order couplings between the condensates and the Polyakov loop, $\sigma^2 \Phi$, $\sigma^3 \Phi$, and $|d|^2 \Phi$ are:

\begin{widetext}
\beqa
a^\prime & = & 144 G^2 \sum_\pm \int \frac{d^3 \mathbf{k}}{(2\pi)^3} \frac{\sinh(\beta E^\pm_0) (f^\pm)^2}{k}   ,   \nonumber \\
c^\prime & = & 216 G K \sum_\pm \int \frac{d^3 \mathbf{k}}{(2\pi)^3} \frac{\sinh(\beta E^\pm_0) (f^\pm)^2}{k}   ,   \\
\alpha^\prime & = & 12 H^2 \sum^\pm \int \frac{d^3 \mathbf{k}}{(2\pi)^3} \bigg \{ \frac{2 \sinh(\beta E^\pm_0) (f^\pm)^2}{E^\pm_0} - 36 \sqrt{3} \pi \beta^2 E^\pm_0 \sinh (\beta E^\pm_0) (h^\pm)^2 g^\pm - 2 \sqrt{3} \pi \beta f^\pm h^\pm    ,   \nonumber \\
	 && \hspace{55mm} + 3 \sqrt{3} \beta \left[2 \sqrt{3} \beta E^\pm_0 \sinh(\beta E^\pm_0) + \pi \right] (f^\pm)^2 h^\pm \bigg \}   .   \nonumber
\eeqa
\end{widetext}

\begin{figure*}
\subfloat{\includegraphics[width=0.45\textwidth,height=0.35\textwidth]{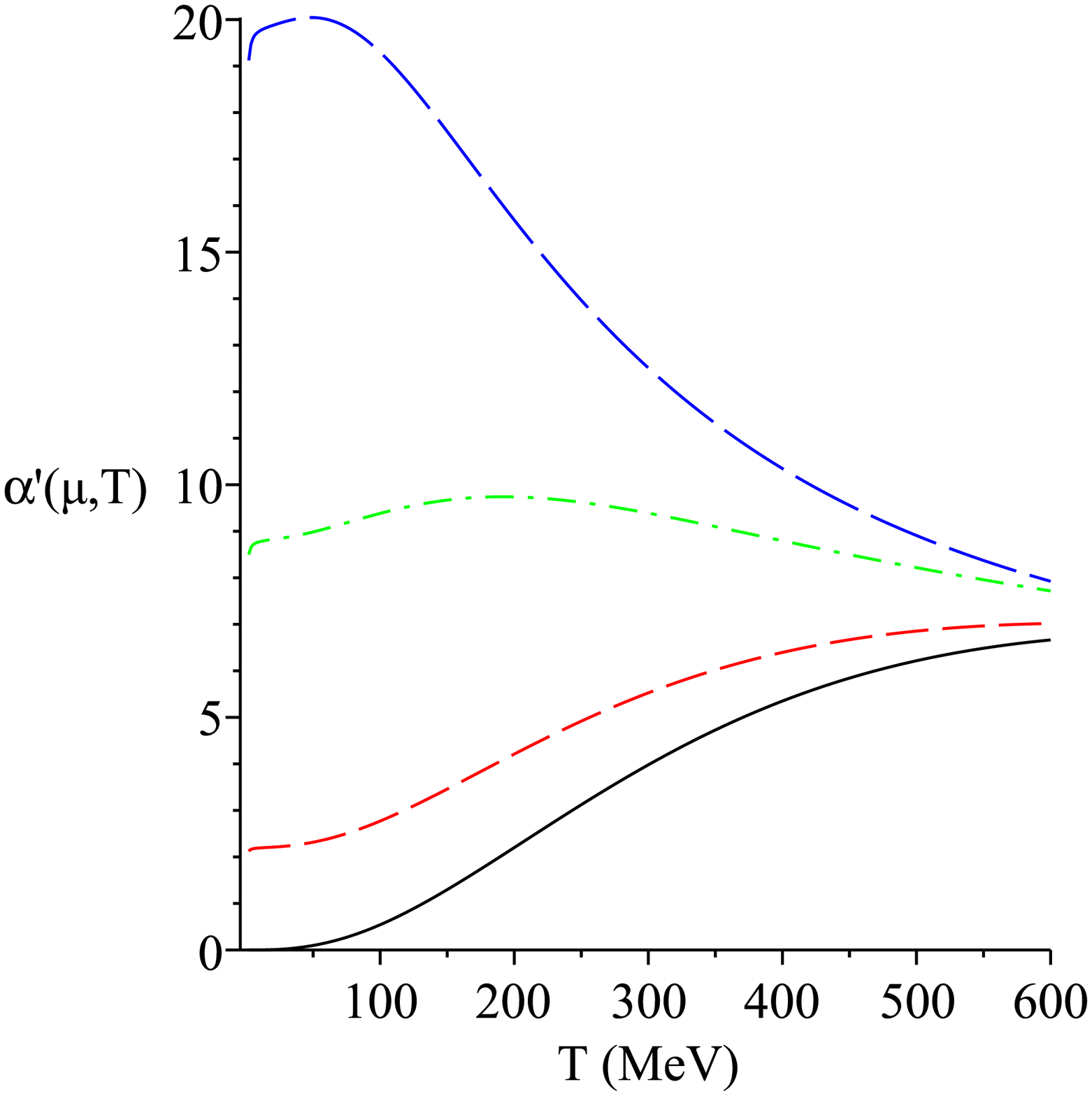} \label{fig:alphap(T)}}
\subfloat{\includegraphics[width=0.45\textwidth,height=0.35\textwidth]{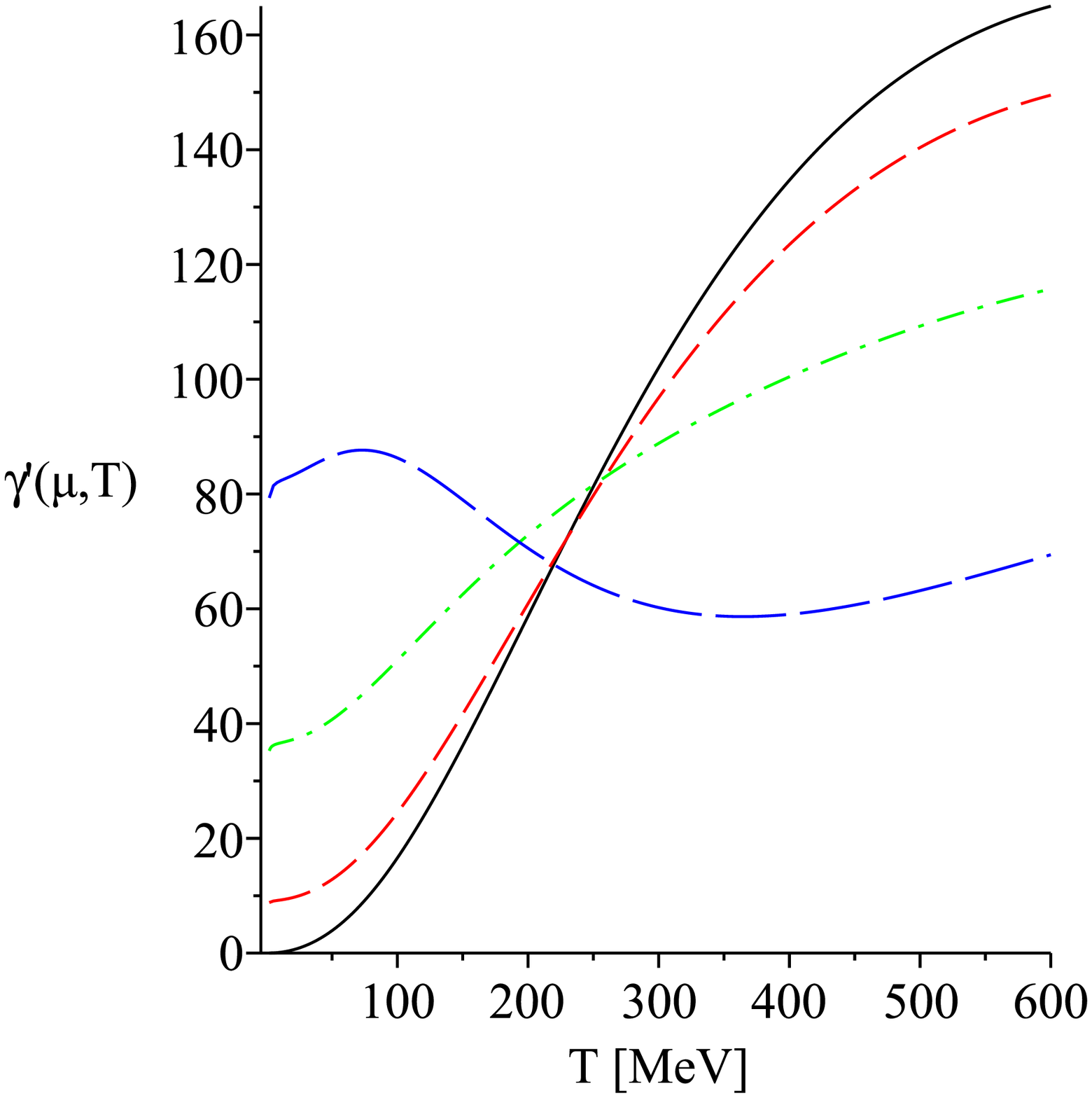} \label{fig:gammap(T)}}
\caption{\footnotesize{(color online) (a) Dimensionless $\alpha^\prime (\mu,T)$ for various values of the chemical potential (same as Fig.~\ref{fig:a(T)}).  For $\alpha^\prime < 0$, the $|d|^2 \Phi$ term favors the coexistence of a diquark condensate and deconfinement.  (b) Coefficient of $\sigma |d|^2 \Phi$, $\gamma^\prime (\mu,T)$, for various values of the chemical potential.  For $\gamma^\prime < 0$, the system favors the coexistence of $\sigma$, $d$, and $\Phi$.}}
\label{fig:alphapgammap(T)}
\end{figure*}

As shown in Fig.~\ref{fig:ap(T)}, except for at very large chemical potential ($\mu \gtrsim 445$ MeV) both $a^\prime$ and $c^\prime = (3 K^\prime / 2 G) a^\prime$ are positive, and therefore tend to disfavor simultaneous chiral condensation and deconfinement.  Since $\sigma$ is only appreciable for $\mu \lesssim 300$ MeV, this means that to lowest-order, the presence of a chiral condensate will tend to maintain a confined state, and vice versa.  We note, however, that this finding does not preclude the realization of a spatially inhomogeneous ``quarkyonic" phase at large $\mu$, as has been suggested recently~\cite{McLerran,Hidaka,Hatsuda3,Carignano,Kojo}, since we have explicitly assumed a homogeneous condensate.

We also note that the magnitude of $a^\prime$ decreases with increasing chemical potential, and is therefore largest (most strongly opposing deconfinement) in the NG region of the phase diagram where $\sigma$ dominates.  This effect is visible in Fig. (\ref{fig:Phicontours}), where the curves of constant $\Phi$ are pulled to higher temperatures in the NG phase by virtue of the presence of a non-zero $\sigma$.  The NG-NOR transition is therefore the result of two competing effects.  On one hand, in the pure-gauge sector confinement tends to become weaker at higher temperatures, eventually giving way to a deconfined QGP.  On the other hand, as temperature increases, the system has an increasing aversion to a state in which both $\sigma$ and $\Phi$ are non-zero, so the presence of the chiral condensate tends to suppress the deconfinement transition.

Similarly, as shown in Fig.~\ref{fig:alphap(T)}, the coefficient $\alpha^\prime$ is positive throughout the phase diagram and its magnitude increases with increasing $\mu$.  Thus, the presence of a diquark condensate also tends to maintain a confined state with $\Phi \sim 0$, and does so more strongly at high chemical potentials, where $d$ is appreciable.  This can be understood by noting that in order to satisfy the gap equation, $\partial \Omega / \partial \Delta = 0$, with increasing $\Phi$, one must decrease the magnitude of the gap.

The final coefficient that we compute is of the term $\sigma |d|^2 \Phi$, the lowest-order term coupling all three variables:

\begin{widetext}
\beqa
\gamma^\prime & = & 6 K^\prime \sum_\pm \int \frac{d^3 \mathbf{k}}{(2\pi)^3} \bigg \{ - 18 \sqrt{3} \pi H \beta^2 E^\pm_0 (h^\pm)^2 \sinh ( \beta E^\pm_0) g^\pm + \left( \frac{3 G}{k} + \frac{H}{E^\pm_0} \right) \sinh (\beta E^\pm_0) (f^\pm)^2   \nonumber \\
	&& \hspace{45mm} - \sqrt{3} \pi H \beta f^\pm h^\pm + \frac{3 \sqrt{3} H \beta}{2} \hspace{.5mm} \left[ 2 \sqrt{3} \beta E^\pm_0 \sinh (\beta E^\pm_0) + \pi \right] (f^\pm)^2 h^\pm \bigg \}   .
\eeqa
\end{widetext}

\noindent As shown in Fig.~\ref{fig:gammap(T)}, $\gamma^\prime$ is positive throughout the phase diagram, so that once again, we find that condensation and deconfinement tend to disfavor one another.  Thus, we find that in general, the existence of a condensate (either $\sigma$ or $d$) tends to encourage confinement, and the greater the number or magnitude of the condensate(s) present, the greater the effect.

\subsection{Low temperature critical point}

Having computed the lowest-order Landau-Ginzburg coefficients, we are now in a position to assess the effect of the Polyakov loop on the low temperature critical point.  As noted in the prior section, in order for the critical point to appear the ratio of the axial anomaly couplings $\kappa = K^\prime / K$ must exceed 4.2.  In the presence of the Polyakov loop, the values of these couplings are effectively modified by terms proportional to $\Phi$, as well as higher-order terms which can be neglected to first-order.  Comparing Eqs. (\ref{eq:V(sigma,d)}) and (\ref{eq:LGomega}) we define
\beqa
K_{eff} & \equiv & K_0 \left(1 + \frac{c^\prime}{c} \hspace{.5mm} \Phi \right)   ,   \\
K^\prime_{eff} & \equiv & K^\prime_0 \left(1 + \frac{\gamma^\prime}{\gamma} \hspace{.5mm} \Phi \right)   . 
\eeqa
Taking the ratio and expanding to linear order in $\Phi$ yields
\beq
\kappa_{eff} = \kappa_0 \left[ 1 + \left(\frac{\gamma^\prime}{\gamma} - \frac{c^\prime}{c} \right) \Phi \right]   .   \label{eq:kappa}
\eeq
The effect of the Polyakov loop can be assessed in terms of the sign of $\delta \equiv \gamma^\prime/\gamma - c^\prime/c$.  For $\delta > 0$, the effective ratio $\kappa_{eff}$ is increased, which will tend to encourage the emergence of the critical point, while for $\delta < 0$ the critical point will tend to be suppressed.  The condition for Eq. (\ref{eq:kappa}) to remain valid is $\delta \ll \Phi^{-1} \sim 2$, which follows from the fact that $\Phi \sim 0.4$ at the critical point.  Noting that for $\kappa_0 = 4.2$ the critical point is at $(\mu,T) = (288 \mbox{ MeV}, 36 \mbox{ MeV})$, we find that $\delta_{cp} = -0.16$.  Thus, the presence of the Polyakov loop decreases the effective coupling ratio $\kappa$ by approximately 6\%.

There are several remaining questions relevant to improving our current picture of the QCD phase diagram.  First, the effects of realistic bare quark masses, particularly the intermediate strange quark mass, on the low temperature critical point have yet to be studied in the context of the PNJL model.  By addressing this problem, we may come to better understand the ``freezing-out" of the strange quark as the system moves to lower chemical potentials, where the light up and down quarks dominate.  Second, we have not imposed the restrictions of charge-neutrality and $\beta$-equilibrium on our analysis.  Finally, our model might be extended by including additional condensate structure beyond CFL, which would allow for a diversity of CSC-COE phases in the low temperature portion of the phase diagram.

\section{Acknowledgements}

This research was supported in part by NSF grant PHY09-69790.  The authors thank Professor Tetsuo Hatsuda for insightful comments throughout the course of this work, and for critically reading the manuscript.   P.P. also thanks Tomoki Ozawa for posing helpful questions and perspective.


\begin{thebibliography}{99}
\bibitem{footnote}There is a slight ambiguity in this definition at low $\mu$.  As shown in Fig.~\ref{fig:mu280}, for $\mu \lesssim 288$ MeV, as the system moves to higher $T$, $\Phi$ first undergoes a smooth, though rapid, crossover, which is followed by a slight discontinuity coincident with the chiral phase transition.  Whether one chooses to identify the deconfinement transition with the maximum finite value of $d \Phi / d T$ (rapid crossover) or with the absolute maximum of $d \Phi / d T$ (discontinuity) is a matter of convention.  We choose the latter definition.  For large $\mu$ there is no such ambiguity, since the COE-NOR transition is second order and no discontinuity exists.
\end{thebibliography}
\end{document}